\documentclass[pra,aps]{revtex4}
\usepackage{graphicx}
\usepackage{epsfig}
\def\ket#1{\left\vert #1 \right\rangle}
\def\bra#1{\left\langle #1 \right\vert}
\def\ensavg#1{\left\langle #1 \right\rangle}
\def\bkm#1#2{\left\langle #1 \vert #2 \right\rangle}
\def\bigo{O}

\def\abs#1{\left\vert #1 \right\vert}

\def\numatom{N}
\def\massatom{m}

\def\freq{\omega}
\def\freqtrap{\omega_0}
\def\freqlight{\omega_c}
\def\freqatom{\omega_a}
\def\ladderup{a^\dagger}
\def\ladderdown{a}

\def\wvecpot{k}
\def\gpot{g}

\def\id{\mathcal{I}}
\def\tensorm{\otimes}

\def\Hint{H_I}
\def\Hfull{H}
\def\Htrap{H_0}
\def\Etrap{E_0}

\def\matpot{V}
\def\matpotop{V}

\def\eps{\epsilon}

\def\v#1{\mathbf{#1}}
\def\matpoteigval{\chi}
\def\matpoteigvec{D}
\def\spaceeigvec{\phi}

\def\toteigvec{\Psi}
\def\totinitvec{\Psi_I}
\def\spaceinitvec{\phi_I}

\def\proj{\Pi}
\def\rotationop{\hat{\mathcal{U}}}

\def\refsec#1{Sec.\ \ref{Section::#1}}

\def\refeqn#1{Eq.\ (\ref{Equation::#1})}

\def\refeqs#1#2{Eqs.\ (\ref{Equation::#1}) and (\ref{Equation::#2})}

\begin{document}

\title{Transmission Spectrum of an Optical Cavity Containing $\numatom$ Atoms}

\author{Sabrina Leslie$^{1,3}$}
\author{Neil Shenvi$^{2,3}$}
\author{Kenneth R. Brown$^{2,3}$}
\author{Dan M. Stamper-Kurn$^{1}$}
\author{K. Birgitta Whaley$^{2,3}$}

\affiliation{Departments of Physics$^1$ and Chemistry$^2$ and the Kenneth
S.
Pitzer Center for Theoretical Chemistry$^3$, University of California,
Berkeley, CA 94720}

\date{\today}

\begin{abstract}
The transmission spectrum of a high-finesse optical cavity
containing an arbitrary number of 
trapped atoms is presented.  We take
spatial and motional effects into account and show that in the
limit of strong coupling, the important spectral features can be
determined for an arbitrary 
number of atoms, $\numatom$.  We also
show that these results have important ramifications in limiting
our ability to determine the number of atoms in the cavity.
\end{abstract}


\maketitle

\section{Introduction} \label{Section::Introduction}

Cavity quantum electrodynamics (CQED) in the strong coupling
regime holds great interest for experimentalists and theorists for
many reasons \cite{berm94book,kimb98,raim01}.  From an applied
perspective, CQED provides precise tools for the fabrication of
devices which generate useful output states of light, as
exemplified by  
the single-photon source
\cite{law97single,kuhn97single,kuhn02single}, the $N$-photon source
\cite{brown03fock}, 
and the optical phase gate \cite{turc95phase}.
Conversely, CQED effects transform the high-finesse cavity
into a sensitive optical detector of objects which are in the
cavity field.  Viewed simply, standard optical microscopy is made
more sensitive by having a probe beam pass through the sample
multiple times, and by efficiently collecting scattered light.  In
the weak-coupling regime, this has allowed for nanometer-
resolution measurements of the positions of a trapped ion
\cite{guth02ion,mundt02}.  In the strong-coupling regime, the
presence and position of single atoms can be detected with high
sensitivity  by monitoring the transmission
\cite{hood00micro,munst99dyn}, phase shift \cite{mabuchi99single},
or spatial mode \cite{horak02kaleid} of probe light sent through
the cavity.

In this paper, we consider using strong-coupling CQED effects to
precisely count the number of atoms trapped inside a high-finesse
optical microcavity.  The principle for such detection is
straightforward: the presence of atoms in the cavity field splits
and shifts the cavity transmission resonance.   A precise $N$-atom
counter could be used to prepare the atoms-cavity system for
generation of optical Fock states of large photon number
\cite{brown03fock}, or to study ultra-cold gaseous atomic systems
gases in which atom number fluctuations are important, such as
number-squeezed \cite{orze01squeeze} and spin-squeezed
\cite{wine92squeeze,hald99,kuzm00qnd} systems.

A crucial issue to address in considering such a CQED device is
the role of the spatial distribution of atoms and their motion in
the cavity field.  An N-atom counter (or any CQED device)  would
be understood trivially if the N atoms to be counted were held at
known, fixed positions in the cavity field.
This is a central
motivation for the integration of CQED with extremely strong traps
for neutral atoms \cite{ye99trap,mckeever03} or ions
\cite{guth02ion,mundt02}. The Tavis-Cummings model \cite{tavis68},
which applies to this case, predicts that the transmission
spectrum of a cavity containing $N$ identically-coupled (with strength $g$), resonant
atoms will be shifted from the empty cavity resonance by a
frequency $g \sqrt{N}$ at low light levels. Atoms in a cavity can
then
be counted by measuring the frequency shift of the maximum cavity
transmission, and 
distinguishing the transmission spectrum of $N$
atoms from that of $N + 1$ atoms in the cavity.  However, to assess the potential for precise CQED-aided probing of
a many-body atomic system, we consider 
here
the possibility that atoms
are confined at length scales comparable to or indeed larger than the
optical wavelength.

In this paper, we characterize the influence of cavity mode
spatial dependence and atomic motion on the transmission spectrum
for an arbitrary number of atoms.  The impact of atomic motion on
CQED has been addressed theoretically in previous work
\cite{ren95,vern97,dohe97motion}, 
although attention has focused primarily on
the simpler problem of a single atom in
the cavity field.  We show that when spatial dependence is
included, the intrinsic limits on atom counting change
significantly.  The organization of this paper as follows. In
\refsec{Transmission} we introduce the system Hamiltonian, define
our notation, and derive an explicit expression for the intrinsic
transmission function.  In \refsec{Moments}, we introduce the
method of moments, and use this method to calculate the shape of
the intrinsic transmission function. Conclusions and implications
for atom counting are presented in \refsec{Conclusions}.

\section{Transmission} \label{Section::Transmission}
Let us consider the Hamiltonian for $\numatom$ identical two-level
atoms in a harmonic potential inside an optical cavity which
admits a single standing wave mode of light.  We consider atomic
motion and the spatial variation of the cavity mode only along the
cavity axis, assuming that the atoms are confined
tightly with respect to the cavity mode waist in the other two
dimensions.  The Hamiltonian for this system is
\begin{equation} \label{Equation::HFull}
\Hfull =\hbar \freqlight \ladderup\ladderdown +
\sum_{i=1}^\numatom{ \hbar\freqatom \ket{e_i}\bra{e_i}} +
\Hfull_{0} + V
\end{equation}
where $\freqlight$ is the frequency of the cavity mode and $\ladderdown(\ladderup)$ is the annihilation (creation) operator for the cavity field. The motional Hamiltonian $\Htrap = \sum_i
\Hfull_{0,i}$ is a sum over single-atom Hamiltonians $\Hfull_{0,i}
= p_i^2/ 2\massatom + \massatom\freqtrap^2 x_i^2/2$ where
$\massatom$ is the atomic mass and $\freqtrap$ the harmonic trap 
frequency.
The atomic ground and excited internal states, $\ket{g}$ and
$\ket{e}$, respectively, are separated by energy $\hbar
\freqatom$.  The dipole interaction with the light field  $V =
\sum_i V_i$ is a sum over interactions with the dipole moment of
each atom $V_i = \hbar\gpot\cos(\wvecpot
x_i)\left(\ket{e_i}\bra{g_i}\ladderdown +
\ket{g_i}\bra{e_i}\ladderup\right)$ where $g$ is the vacuum-Rabi
splitting, which depends on the atomic dipole moment and the volume
of the cavity mode. In this paper we assume the cavity mode frequency to be in exact resonance with the
atomic transition frequency, $\freqlight = \freqatom$.

Since the Hamiltonian (Eq.\ \ref{Equation::HFull}) commutes with
the total excitation operator, $n_T = \ladderup\ladderdown +
\sum_i {\ket{e_i}\bra{e_i}}$, the eigenspectrum of $\Hfull$ breaks
up into manifolds labelled by their total excitation number.  In
this work, we are concerned with excitation spectra of the
atoms-cavity system at the limit of low light intensity, and we
therefore restrict our treatment to the lowest two manifolds, with
$n_T = \{0,1\}$.

In particular, we consider the excitation spectra from the ground
state (motional and internal) of the atoms-cavity system.  This
state $\ket{\Psi_0}$ is given simply as a product of motional and
internal states, $\ket{\Psi_0} = \ket{\Phi_I} \tensorm |0_c;g_1,
g_2, \ldots g_N\rangle$. In the uncoupled internal state
notation, the $0_c$ symbol indicates there are zero photons in the
cavity, and the $g_i$ symbol indicates that atom $i$ is in the
ground state.The motional state $\ket{\Phi_I} =
\prod_{i=1}^\numatom{\ket{\phi_0(x_i)}}$ is a product of
single-atom ground states of the harmonic trap.

Let us calculate the low-light intensity transmission spectrum of
the cavity.   We  assume that the system is pumped by a
near-resonant linearly coupled driving field such that the cavity
excitation Hamiltonian is $ \Hint = E\left(\ladderup e^{-i\freq
t} + \ladderdown e^{i\freq t}\right)$ where $E$ is the product of
the external driving electric field strength and the
transmissivity of the input cavity mirror, and $\freq$ is the
driving frequency. To determine the cavity transmission spectrum,
we determine the excitation rate to atoms-cavity states in the
$n_T = 1$ manifold from the initial ground state.  The
atoms-cavity eigenstates decay either by cavity emission, with the
transmitted optical power proportional to $\kappa \ensavg{N_c}$
where $\kappa$ is the cavity decay rate and $N_c =
\ladderup\ladderdown$ is the intracavity photon number operator,
or by other processes (spontaneous emission, losses at the
mirrors, etc.) at the phenomenological rate constant $\gamma$.
Neglecting the width of the transmission spectrum caused by cavity
and atomic decay ($\kappa, \gamma \to 0$), we  use Fermi's Golden
Rule to obtain the transmission spectrum $I(\freq)$:
\begin{equation} \label{Equation::IOmega}
I(\freq) \propto \sum_{j,n_{T}=1}{
\abs{\bra{\Psi_j}\ladderup\ket{\Psi_0}}^2\delta(\freq_j-\freq_0 -
\freq)} = \sum_{j,n_{T}=1}{
\abs{\bkm{\Psi_{j}}{\Psi_I}}^2\delta(\freq_{j}-\freq_0 - \freq)},
\end{equation}
where $\ket{\Psi_I} = \ladderup\ket{\Psi_0}$. In the summation over all atoms-cavity eigenstates, we make the simplification that only states with $n_T=1$ need be included since only these states are coupled to the ground state by a single excitation. To simplify notation, we make this implicit assumption throughout the remainder of this paper. We denote $I(\freq)$ as the
``intrinsic transmission spectrum''.  In the limit of $\kappa,
\gamma \to 0$ this is composed of delta functions in frequency, 
while an experimentally
observed transmission spectrum would be convolved by non-zero
linewidths.

To proceed further, we introduce the basis states $\{\ket{0};
\ket{i}\}$ which span the space of \emph{internal states} in the
$n_T = 1$ manifold. 
The state $\ket{0} = \ket{1_c; g_0, g_1,
\ldots g_N}$ has one cavity photon and all atoms in their
ground state. 
The state $\ket{i} = \ket{0_c; g_0, g_1,
\ldots e_i \ldots g_N}$ is the state in which the cavity field is
empty, while a single atom (atom $i$) is in the excited state.
Restricted to the $n_T = 1$ manifold, the Hamiltonian (Eq.\
(\ref{Equation::HFull})) is written as $\Hfull = \Htrap + V_{n_T =
1}$, where
\begin{equation} \label{Equation::HNoTrap}
\matpotop_{n_T = 1} = \sum_i \hbar\gpot\cos{\left(\wvecpot
x_i\right)}\tensorm\left(\ket{i}\bra{0}+\ket{0}\bra{i}\right).
\end{equation}

To gain intuition regarding the behavior of the system, let us
define the operator $\matpot(\v{x})$ as the optical potential
operator $V_{n_T = 1}$ for which the position operators are
replaced by definite positions $\v{x}$.  In the $N+1$ dimensional
space of internal states for the $n_T = 1$ manifold, the operator $\matpot(\v{x})$ has two non-zero eigenvalues,
$\pm\hbar\gpot\matpoteigval(\v{x}) = \pm 
\hbar\gpot\sqrt{\sum_i{\cos^2{\wvecpot
x_i}}}$ with corresponding eigenstates
\begin{equation} \label{Equation::MatPotEigVec}
\ket{\matpoteigvec_{\pm}(\v{x})} = \frac{1}{\sqrt{2}} \left(
\ket{0} \pm \frac{1}{\matpoteigval(\v{x}) } \sum_i \cos{\wvecpot
x_i} \ket{i}\right).
\end{equation}
We will refer to the $\ket{\matpoteigvec_-(\v{x})}$ and
$\ket{\matpoteigvec_+(\v{x})}$ eigenstates of the potential matrix
as the red and blue internal states, respectively, in reference to
their energies being red- or blue-detuned from the empty cavity
resonance. The remaining $N-1$ eigenvalues of the optical potential
matrix are 
null-valued.  These
correspond to dark states having no overlap with
the excited cavity internal state, $\ket{0}$, and which,
therefore, cannot be excited by the cavity excitation interaction
$\Hint$.  Note that $\ensavg{N_c}=1/2 (0)$ for all bright (dark) states, hence the cavity transmission spectrum is equivalent to the excitation spectrum in this treatment. We can now write the optical potential operator
$\matpotop_{n_T = 1}$ as
\begin{equation}
\matpotop_{n_T = 1}= \gpot\int{d\v{x} \,
 \chi(\v{x}) \, \ket{\v{x}}\bra{\v{x}}
\tensorm \left( \ket{\matpoteigvec_+(\v{x})}
\bra{\matpoteigvec_+(\v{x})} - \ket{\matpoteigvec_-(\v{x})}
\bra{\matpoteigvec_-(\v{x})} \right)}.
\end{equation}
We also note that the initial state $\ket{\totinitvec}$ can be written as 
a superposition of bright states,
\begin{equation}\label{TotInitVecDpm}
\ket{\totinitvec} = 
\frac{1}{\sqrt{2}}\left(
\ket{\spaceinitvec(\v{x})\tensorm\matpoteigvec_-(\v{x})} + 
\ket{\spaceinitvec(\v{x})\tensorm\matpoteigvec_+(\v{x})}\right).
\end{equation}

Our treatment allows us to recover easily results of the
Tavis-Cummings model \cite{tavis68} in which a collection of fixed
two-level atoms are coupled to a single-mode cavity with fixed,
identical dipole coupling.  Considering $\matpotop(\v{x_0})$ with
all atoms at the origin ($\v{x_0} = \left(0, 0, \ldots 0\right)$),
we find a spectrum composed of delta-functions at $\pm
\gpot\sqrt{\numatom}$ (see Figure 1a) corresponding to the two
bright states $\ket{\matpoteigvec_\pm(\v{x_0})}$.  The clear
dependence of the frequency of peak transmission on the integer
number of atoms in the cavity provides the background for a basic,
transmission-based atom-counting scheme. ``Extrinsic''
line-broadening, due to cavity decay and other losses, will smear
out these sharp transmission peaks (see Figure 1b), and will determine the 
maximum number of atoms that can be counted at the single-atom level by
discriminating between the transmission spectra for $N$ and $N+1$
atoms. For the remainder of the paper, we focus on intrinsic limitations
to atom counting, i.e.\ those due to atomic localization and
motion.

\section{Method of Moments} \label{Section::Moments}

To analyze the transmission characteristics of the atoms-cavity
system in the presence of spatial dependence and atomic motion, we
shall assume that the key features of the spatially-independent limit
discussed above are maintained (Figure 2).  Specifically, the
transmission spectrum will still be described by two sidebands,
one red-shifted and one blue-shifted from the empty cavity resonance by
some frequency on the order of $\gpot$.  In determining the cavity
transmission $I(\freq)$, we may thus
divide the bright excited states $\{ \ket{\Psi_{j}}\}$ of the $n_T =
1$ manifold into ``red'' $\{ \ket{\Psi_{j,-}}\}$ and ``blue'' $\{ \ket{\Psi_{j,+}}\}$ states. From these  ``red'' and  ``blue'' states, we determine the transmission lineshapes $I_-(\freq)$ and $I_+(\freq)$ of the red and blue sidebands, respectively.

The validity of this approach is made more exact by the following
considerations.  We have already obtained the locally-defined
internal-state eigenbasis for the $n_T = 1$ manifold as
eigenstates of the operator $\matpot(\v{x})$, namely the states
$\ket{\matpoteigvec_\pm(\v{x})}$ and the remaining $N-1$ dark
states. Let $\rotationop(\v{x})$ be the rotation operator which
connects the uncoupled internal states $\{\ket{0}, \ket{1}, \dots
\ket{N}\}$ to the eigenstates of $\matpot(\v{x})$ at a particular
set of coordinates $\v{x}$ (the ``coupled internal state basis'').
Now, consider applying this local choice of ``gauge'' everywhere
in the system. Since the dipole interaction operator $\matpotop$ is
diagonalized in the coupled internal state basis, it is convenient
to examine the full Hamiltonian $\Hfull$ in this basis. Defining
the spatially-dependent rotation operator $\rotationop =
\int{d\v{x} \, \ket{\v{x}}\bra{\v{x}} \, \rotationop(\v{x})}$, we
therefore consider the transformed Hamiltonian $\Hfull^\prime =
\rotationop \Hfull \rotationop^\dagger$.

Returning to \refeqn{HFull}, the only portion of the Hamiltonian
$\Hfull$ which does not commute with the operator $\rotationop$ is
the kinetic energy. Considering the transformation of the momentum
operator for atom $i$,

\begin{equation}
\rotationop p_i \rotationop^\dagger = p_i + \frac{\hbar}{i}
\rotationop \frac{d}{d x_i} \rotationop^\dagger = p_i + A_i
\end{equation}
the transformed Hamiltonian $\Hfull'$ can be expressed as $\Hfull' = \Hfull_{ad} + \Delta \Hfull$, where

\begin{eqnarray}
H_{ad} &=& \sum_i{\left(\frac{p_i^2}{2\massatom}\tensorm\id + 
\frac{1}{2}\massatom\freqtrap^2x_i^2\tensorm\id\right)}
+\hbar\gpot\chi(\v{x})\left(\ket{D_+}\bra{D_+}-\ket{D_-}\bra{D_-}\right), \\
\Delta \Hfull &=& \frac{1}{2\massatom}\sum_i{\left( p_i A_i + A_i p_i + A_i 
A_i\right)}.
\label{Equation::DeltaH}
\end{eqnarray}
The operator $\Hfull_{ad}$ describes the 
behavior of atoms which adiabatically follow the coupled internal state 
basis while moving through the spatially-varying cavity field, 
and 
$\Delta \Hfull$ represents the kinetic energy associated with this local 
gauge definition.  

Let us treat  $\Delta H$ as a perturbation and expand the eigenvalues and eigenstates of  $\Hfull'$
as
\begin{eqnarray}
E_{j,\pm} &=& E_{j,\pm}^{(0)} + E_{j,\pm}^{(1)} + \ldots,\\
\ket{\toteigvec_{j,\pm}} &=& \ket{\toteigvec_{j,\pm}^{(0)}} + 
\ket{\toteigvec_{j,\pm}^{(1)}} + \ldots
\end{eqnarray}
We define projection operators onto the red 
and blue and dark internal states, $\proj_-, \proj_+, \proj_d$, respectively, with the explicit forms  
\begin{equation}
\proj_\pm = \int{d\v{x} \ket{\v{x}}\bra{\v{x}}
\tensorm\ket{\matpoteigvec_\pm(\v{x})}\bra{\matpoteigvec_\pm(\v{x})}}.
\end{equation}
These projection operators commute with $\Hfull_{ad}$.
Hence the bright eigenstates of $\Hfull_{ad}$, which are simultaneous eigenstates of $\proj_\pm$ and $\proj_d$, can be written 
as 
\begin{equation}
\ket{\toteigvec_{j,\pm}^{(0)}} = \ket{\spaceeigvec_{j,\pm}^{(0)} \tensorm
\matpoteigvec_\pm} \equiv \int{ d\v{x} \,
\spaceeigvec_{j,\pm}^{(0)}(\v{x}) \, \ket{\v{x}} \tensorm
\ket{\matpoteigvec_\pm(\v{x})}}.
\end{equation}

We now assign an eigenstate, $\ket{\toteigvec_j}$, of $H'$ 
to the red or blue sideband if its zeroth order component 
$\ket{\toteigvec_j^{(0)}}$ belongs respectively to the 
$\ket{\matpoteigvec_-}$ or $\ket{\matpoteigvec_+}$ 
manifold.  We can therefore define the sideband transmission spectra 
$I_\pm(\freq)$ as the separate contributions of red/blue sideband states
to the total transmission spectra (see \refeqn{IOmega}):
\begin{equation} \label{Equation::IOmega_pm}
I_\pm(\freq) \propto \sum_j{ \abs{\bkm{\toteigvec_{j,\pm}}{\totinitvec}}^2
\delta(\freq_{j,\pm} -\freq_0 - \freq)}.
\end{equation}
Determining the exact form of  $I_\pm(\freq)$ is equivalent to
solving for all the eigenvalues $\hbar \freq_{j,\pm}$ of the full Hamiltonian.
This is a difficult problem, particularly as the
number of atoms in the cavity increases.  In practice, given the
potential extrinsic line-broadening effects which may preclude
the resolution of individual
spectral lines, it may suffice to simply
characterize main features of the transmission spectra.  As we
show below, general expressions for the various moments of the
spectral line can be obtained readily as a perturbation expansion in 
$\Delta H$.  These moments allow one to assess the feasibility of 
precisely counting the number of atoms contained in the high-finesse 
cavity based on the transmission spectrum.

In general, we evaluate averages $\ensavg{\freq_\pm^n}$ weighted
by the transmission spectral distributions $I_\pm(\freq)$.  We
make use of the straightforward identification (for notational
clarity, shown here explicitly for the case of the blue sideband)
\begin{eqnarray} \label{Equation::EnsAvgE}
\hbar\ensavg{\freq_+} &=& \frac{\hbar\int{ d\freq \, I_+(\freq) \, 
\freq}}{\int{d\freq \, I_+(\freq)}} \\
&=& 
\frac{\sum_{j}{E_{j,+}\bkm{\totinitvec}{\toteigvec_{j,+}}\bkm{\toteigvec_{j,+}}{\totinitvec}}}
{\sum_{j}{\bkm{\totinitvec}{\toteigvec_{j,+}}\bkm{\toteigvec_{j,+}}{\totinitvec}}}\\
&=&
\frac{\sum_{j}{E_{j,+}\bra{\totinitvec}
\left(\proj_+ + \proj_-\right)
\ket{\toteigvec_{j,+}}\bra{\toteigvec_{j,+}}
\left(\proj_+ + \proj_-\right)
\ket{\totinitvec}}}
{\sum_{j}{\bra{\totinitvec}
\left(\proj_+ + \proj_-\right)
\ket{\toteigvec_{j,+}}\bra{\toteigvec_{j,+}}
\left(\proj_+ + \proj_-\right)
\ket{\totinitvec}}},
\end{eqnarray}
where we have made use of the facts that $\proj_+ + \proj_- + \proj_d 
= I$ and $\proj_d \ket{\totinitvec} = 0$.  To zeroth 
order, 
\refeqn{EnsAvgE} becomes,
\begin{eqnarray}
\hbar\ensavg{\omega_+}^{(0)} &=&
\frac{\sum_{j}{E_{j,+}^{(0)}\bkm{\totinitvec}{\toteigvec_{j,+}^{(0)}}\bkm{\toteigvec_{j,+}^{(0)}}{\totinitvec}}}
{\sum_{j}{\bkm{\totinitvec}{\toteigvec_{j,+}^{(0)}}\bkm{\toteigvec_{j,+}^{(0)}}{\totinitvec}}} =
 2\bra{\totinitvec}{\proj_+}H_{ad}{\proj_+}\ket{\totinitvec}. 
\end{eqnarray}
The first-order correction to this result is given by,
\begin{equation}
\begin{array}{rcl}
\hbar\ensavg{\omega_+}^{(1)} &=& 2\left(\bra{\totinitvec}\proj_+\Delta 
H\proj_+\ket{\totinitvec} + 
\bra{\totinitvec}\proj_-\sum_j{E_{j,+}^{(0)}\ket{\toteigvec_{j,+}^{(1)}}
\bra{\toteigvec_{j,+}^{(0)}}}\proj_+\ket{\totinitvec} +
\bra{\totinitvec}\proj_+\sum_j{E_{j,+}^{(0)}\ket{\toteigvec_{j,+}^{(0)}}
\bra{\toteigvec_{j,+}^{(1)}}}\proj_-\ket{\totinitvec}
\right)\\
&& - 4 
\bra{\totinitvec}\proj_+H_{ad}\proj_+\ket{\totinitvec}
\left(\bra{\totinitvec}\proj_-\sum_j{\ket{\toteigvec_{j,+}^{(1)}}
\bra{\toteigvec_{j,+}^{(0)}}}\proj_+\ket{\totinitvec} +
\bra{\totinitvec}\proj_+\sum_j{\ket{\toteigvec_{j,+}^{(0)}}
\bra{\toteigvec_{j,+}^{(1)}}}\proj_-\ket{\totinitvec}
\right).
\end{array}
\end{equation}

To evaluate the sums over the first-order corrections to the 
eigenstates, $\ket{\toteigvec_{j,\pm}^{(1)}}$, we approximate the energy 
denominator in the first-order perturbation correction as the difference between the average energies of the red and blue sidebands,
\begin{eqnarray}
\bra{\totinitvec}\proj_-\sum_j{\ket{\toteigvec_{j,+}^{(1)}}
\bra{\toteigvec_{j,+}^{(0)}}}\proj_+\ket{\totinitvec}
&=&
\bra{\totinitvec}\sum_j{\sum_k{
\frac{\ket{\toteigvec_{k,-}^{(0)}}\bra{\toteigvec_{k,-}^{(0)}}\Delta 
H\ket{\toteigvec_{j,+}^{(0)}}}{\hbar\omega_{j,+}^{(0)} - \hbar\omega_{k,-}^{(0)}}
\bra{\toteigvec_{j,+}^{(0)}}}}\proj_+\ket{\totinitvec},\\
&\approx&
\frac{1}{\ensavg{\hbar\omega_+^{(0)}} - 
\ensavg{\hbar\omega_-^{(0)}}}\bra{\totinitvec}\proj_-\Delta H\proj_+\ket{\totinitvec}.
\end{eqnarray}
Using this approximation, we can evaluate \refeqn{EnsAvgE} to
the first order in perturbation, yielding
\begin{equation} \label{Equation::EnsAvgOmega}
\begin{array}{ll}
\hbar\ensavg{\omega_+} = &2\ensavg{\proj_+ H \proj_+} + 
\frac{1}{\ensavg{\hbar\omega_+^{(0)}}-\ensavg{\hbar\omega_-^{(0)}}}\bigl(
4\ensavg{\proj_+H_{ad}\Delta H \proj_-}  
- 8\ensavg{\proj_+ H_{ad}\proj_+}\ensavg{\proj_- \Delta H \proj_+}
\bigr),
\end{array}
\end{equation}
where all expectation values are calculated over the initial state 
$\ket{\totinitvec}$.  We can also calculate the second moment of the 
distribution using the same technique.  To first-order, we obtain,
\begin{equation} \label{Equation::EnsAvgOmega2}
\begin{array}{ll}
\hbar^2\ensavg{\omega_+^2} = &2\ensavg{\proj_+ \left(H_{ad}^2 + \Delta H 
H_{ad} + H_{ad} \Delta H\right) \proj_+} + 
\frac{1}{\ensavg{\hbar\omega_+^{(0)}}-\ensavg{\hbar\omega_-^{(0)}}}\bigl(
4\ensavg{\proj_+H_{ad}^2\Delta H \proj_-} \\
&- 8\ensavg{\proj_+ H_{ad}^2\proj_+}\ensavg{\proj_- \Delta H \proj_+}
\bigr).
\end{array}
\end{equation}

In order to evaluate these expressions, we must calculate 
expectation values of the form $\proj^\pm H_{ad}^j \Delta H^k \proj^\pm$ 
over the initial state $\ket{\totinitvec}$.  To simplify matters, we note 
that we can 
act with the projection operators on the initial state $\ket{\totinitvec}$,  
which is equivalent to operating in the $\ket{D_\pm}$ internal state 
basis.  
Since $H_{ad}$ is diagonal in the $\ket{D_\pm}$ basis, and 
 $\spaceinitvec(\v{x})$ is the $\numatom$-dimensional harmonic 
oscillator ground state, it 
is straightforward to obtain
\begin{equation} \label{Equation::HadTotInitVec}
H_{ad} \ket{\spaceinitvec(\v{x})D_\pm} = \left(E_0 + 
\hbar \gpot \chi(\v{x})\right) \ket{\spaceinitvec(\v{x})D_\pm}.
\end{equation}
Using the definition in \refeqn{DeltaH}, we find that the $\ket{D_\pm}$ 
matrix elements of $\Delta H$ are given by the matrix
\begin{equation} \label{Equation::DeltaHDpm}
\Delta H = \frac{\hbar^2\wvecpot^2\zeta(\v{x})}{4\massatom}\tensorm\left(
\begin{array}{cc}
1 & -1 \\
-1 & 1
\end{array}
\right),
\end{equation}
where we have defined,
\begin{equation}
\zeta(\v{x}) = -\frac{N-1}{\chi^2} + 1 - 
\sum_{i=1}^N{\frac{\cos^4{(\wvecpot x_i)}}{\chi^4}}.
\end{equation}
Combining \refeqs{EnsAvgOmega}{EnsAvgOmega2} with 
\refeqs{HadTotInitVec}{DeltaHDpm}, we obtain to first-order in $\Delta H$,
\begin{eqnarray}\label{Equation::EnsAvgOmegaLambda}
\hbar\ensavg{\omega_+} - E_0&=& \hbar\gpot\ensavg{\chi} + 
\frac{1}{2}\frac{\hbar^2\wvecpot^2}{2\massatom}\ensavg{\zeta} + 
\frac{1}{2}\frac{\hbar^2\wvecpot^2}{2\massatom}\frac{1}{\ensavg{\chi}}\left(-\ensavg{\zeta 
\chi } + \ensavg{\zeta}\ensavg{\chi}\right), \\
\label{Equation::EnsAvgOmega2Lambda}
\hbar^2\left(\ensavg{\omega_+^2}-\ensavg{\omega_+}^2\right) &=& 
\hbar^2\gpot^2\left(\ensavg{\chi^2}-\ensavg{\chi}^2\right) + 
\hbar\gpot\frac{\hbar^2\wvecpot^2}{2\massatom}
\left(\frac{1}{\ensavg{\chi}}\left(\ensavg{\zeta \chi^2} + 
\ensavg{\zeta}\ensavg{\chi^2}\right)-2 \ensavg{\zeta}\ensavg{\chi}\right).
\end{eqnarray}
Here all expectation values are taken over the spatial state 
$\spaceinitvec(\v{x})$. Although the function $\spaceinitvec(\v{x})$ is simply the product of $\numatom$ harmonic oscillator ground states, the presence of various powers of $\matpoteigval(\v{x})$ and $\zeta(\v{x})$ in the above expectation values makes their analytic evaluation very difficult for arbitrary 
$\numatom$.To determine the dependence of these integrals on atom
number $\numatom$, one may expand the integrand as a Taylor series
in $\matpoteigval^2$, leading to approximate analytic solutions
for the integral as a series in $1/\numatom$.  After some tedious
algebra, we find the average positions of the red- and blue-transmission sidebands to be
\begin{equation} \label{Equation::ERedBlue}
\begin{array}{rcl}
\hbar\ensavg{\omega_\pm}- E_0 &=& \pm 
\hbar\gpot\sqrt{\numatom}\sqrt{\frac{1+\eps}{2}}
\left(1-\frac{1}{\numatom}\frac{(1-\eps)^2}{16}\right)
-\frac{\hbar^2\wvecpot^2}{2\massatom}\left(\frac{1-\eps}{2(1+\eps)}\right)
+\bigo\left(\frac{1}{N}\right).
\end{array}
\end{equation}
Here we quantify the relative length scales of the initial harmonic
trap as compared to the optical interaction potential through the 
parameter $\epsilon = \exp{\left(-\wvecpot^2
\sigma^2 \right)}$, 
which is
related to the Lamb-Dicke parameter $\eta$ by $\sqrt{2}\eta = \wvecpot\sigma$ and $\sigma = \sqrt{\hbar / m \freqtrap}$.

Next, we obtain an expression for the width of the red and
blue sidebands by evaluating the second moment of the sidebands. Expanding \refeqn{EnsAvgOmega2Lambda} as 
a series in $1/\numatom$, we 
obtain
\begin{equation} \label{Equation::WidthRedBlue}
\begin{array}{rcl}
\hbar^2\left(\ensavg{\omega^2_\pm} - \ensavg{\omega_\pm}^2\right) &=&
\frac{1}{16}\hbar^2\gpot^2\left(1-\eps\right)^2\left(1+\eps\right)
\pm\hbar\gpot\frac{\hbar^2\wvecpot^2}{2\massatom}\frac{1}{4\sqrt{\numatom}\sqrt{2\left(1+\eps\right)}}\left(1-\eps\right)^2\left(3+\eps\right)
+\bigo\left(\frac{1}{\numatom}\right).
\end{array}
\end{equation}

To gain some physical insight into these results, we consider two
important regimes: the tight and loose trap regimes.   These 
different
regimes are reflected in the corresponding values of the parameter 
$\epsilon$, which tends
towards $1$ in the extreme tight-trap limit and to $0$ in the extreme
loose-trap limit.  In the tight regime, the length scale of the trapping potential is much
smaller than the wavelength of the light, i.e. $\wvecpot \sigma
\ll 1$.
This is
 equivalent to the Lamb-Dicke regime and is applicable to
current experiments for trapped ions in cavities
\cite{guth02ion,mundt02}, or for neutral atoms held in deep optical
potentials \cite{ye99trap}. In the loose-trap regime, $\wvecpot
\sigma \geq 1$ and atoms in the ground state of the harmonic
oscillator potential are spread out over a distance comparable to
the optical wavelength. As atoms in this regime sample broadly the
cavity field, one expects, and indeed finds, a significant
inhomogeneous broadening of the atoms-cavity resonance.

In the extreme loose-trap limit ($\epsilon \to 0$), we find
\begin{eqnarray}
\ensavg{\freq_\pm} - \Etrap/\hbar  &=& \pm \gpot
\sqrt{\frac{\numatom}{2}}\left(1-\frac{1}{16\numatom}\right)+
\frac{1}{2}\frac{\hbar \wvecpot^2}{2\massatom} + 
\bigo\left(\frac{1}{\numatom}\right), \\
\ensavg{(\Delta \omega_\pm)^2} &=& \frac{1}{8}\gpot^2 \pm \gpot 
\frac{\hbar\wvecpot^2}{2\massatom}\frac{3}{4\sqrt{2\numatom}} + 
\bigo\left(\frac{1}{\numatom}\right).
\end{eqnarray}
In the loose-trap limit, the center of the red sideband is now located at
$\gpot\sqrt{\numatom/2}$ instead of at $\gpot\sqrt{\numatom}$ 
as we
obtained for the spatially independent case.  This difference is due to
the spatial dependence of the standing mode; the atoms no longer always
feel the full strength of the potential, but 
are sometimes located at
nodes of the potential.  We also see that the sidebands have an intrinsic
width of $\approx \gpot/\sqrt{8}$.  This width will play an important part
in limiting our ability to count the number of atoms in the cavity
in the limit of a loose trap. 

Considering the tight-trap limit, we expand in the small parameter
$\wvecpot \sigma$ and obtain
\begin{eqnarray}
\ensavg{\freq_\pm} - \Etrap/\hbar &=& \pm \gpot
\sqrt{\numatom}\left(1-\frac{1}{4}\wvecpot^2\sigma^2\right)
-\frac{1}{4}\frac{\hbar 
\wvecpot^2}{2\massatom}\wvecpot^2\sigma^2+\bigo\left(\wvecpot^4\sigma^4\right),
\\
\ensavg{(\Delta \omega_\pm)^2} &=& 
\frac{1}{8}\gpot^2\wvecpot^4\sigma^4\pm\gpot\frac{\hbar\wvecpot^2}{2\massatom}\frac{1}{2\sqrt{\numatom}}\wvecpot^4\sigma^4 
+ \bigo\left(\wvecpot^6\sigma^6\right).
\end{eqnarray}
In the limit $\wvecpot \sigma \to 0$, the atoms are confined to
the origin and we recover the Tavis-Cummings result discussed
earlier, wherein the transmission sidebands are delta functions at $\pm g
\sqrt{\numatom}$ away from the empty cavity resonance.  As the tightness 
of the trap decreases, the atoms begin to expericence the weaker regions 
of the optical potential and the centers of the sidebands move towards the 
origin.  In addition, the sidebands develop an intrinsic variance which 
scales as $\wvecpot^4\sigma^4$.  

An important feature of both regimes is the intrinsic linewidth
of both the red and blue sidebands (see Figure 2a). This linewidth has a 
magnitude of approximately $\gpot\sqrt{(1-\epsilon)/8}$ when the vacuum Rabi splitting  is much larger than the atomic recoil
energy, i.e., $\gpot \gg \hbar \wvecpot^2 / 2 m$. It is unrelated to linewidth due to cavity decay or spontaneous emission 
which we have 
not addressed here and results purely from the spatial dependence of
the atom-cavity coupling.  Thus, it will provide an intrinsic
limit to our ability to count $\numatom$ 
atoms, regardless of the quality of the 
cavity that is used. 
Our expression for the intrinsic linewidth also highlights an asymmetry 
between the red and blue sidebands. To first-order, increasing the atomic 
recoil energy {\it reduces} the linewidth of the red sideband but increases the 
linewidth of the blue sideband. Consequently probing the red sideband of the 
atoms-cavity system rather than the blue sideband would facilitate counting atoms. In addition, these results suggest that the ability to tune both the atomic 
recoil 
energy $ \hbar \wvecpot^2 / 2 m$ and the coupling strength $g$ 
(this can be done, for instance, using CQED on Raman transitions) 
would be beneficial. We attribute the asymmetry between the sidebands to the 
different effective potentials seen by states within the red and blue 
sidebands. A detailed analysis of this aspect will be provided in a future 
publication.

\section{Conclusions} \label{Section::Conclusions}

We have found that the transmission spectrum of the cavity
containing $\numatom$ atoms trapped initially in the ground state of an
harmonic potential will consist of distinct transmission sidebands
which are red- and blue-detuned from the bare-cavity resonance,
when the vacuum Rabi splitting dominates the
atomic recoil energy.
Analytic expressions for the first and second moments of the
transmission sidebands were derived, and evaluated in the limits
of tight and loose initial confinement.  These expressions include
terms
 containing the vacuum Rabi splitting $\hbar\gpot$ and the
recoil energy $\hbar^2 \wvecpot^2 / 2 m$.  The former can be
regarded as line shifts and broadenings obtained by quantifying
inhomogeneous broadening under a local-density approximation,
i.e.\ treating the initial atomic state as a statistical
distribution of infinitely massive atoms. The latter quantifies
residual effects of atomic motion, in essence quantifying 
effects of
 Doppler
shifts and line broadenings.

These results can be applied to assess the potential for precisely
counting the number of atoms trapped in a high-finesse optical
cavity through measuring the transmission of probe light, 
analogous
to the work of Hood \emph{et al.}\ \cite{hood00micro} and M\"{u}nstermann \emph{et
al.}\ \cite{hood00micro,munst99dyn}
for single atom detection.  To set the limits of our
counting capability, we assume that atoms are detected through
measuring the position of the mean of the red sideband. In order
to reliably distinguish between $\numatom$ and $\numatom+1$ atoms
in the cavity, the difference between the means for $\numatom$ and
$\numatom + 1$ atoms must be greater than the width of our peaks, i.e., 
$\abs{\ensavg{\freq_\pm(\numatom)} -
\ensavg{\freq_\pm(\numatom+1)}} > \Delta \omega_\pm $ (see Figure 3).
Let us consider that, in addition to the intrinsic broadening
derived in this paper, there exists an extrinsic width
$\kappa^\prime$ due to the finite cavity finesse and other
broadening mechanisms.  Evaluated in the limit $\gpot \gg \hbar
\wvecpot^2 / 2 m$ and assuming large $N$,
\begin{equation}
\ensavg{\freq_-(\numatom)} - \ensavg{\freq_-(\numatom+1)} \simeq
\gpot \sqrt{\frac{1+\epsilon}{8\numatom}}.
\end{equation}
We thus obtain an atom counting limit of
\begin{equation}
\numatom_{max} \simeq \frac{1+\epsilon}{8\frac{\kappa^{\prime \,
2}}{\gpot^2}+\frac{1}{2}(1-\epsilon)^2(1+\eps)}.
\end{equation}
where we have assumed that the intrinsic and extrinsic widths add in
quadrature.  This atom counting limit ranges from $\numatom_{max}
= g^2 / 4 \kappa^{\prime \, 2}$ in the tight-trap limit, to
$\numatom_{max} = 1/(1/2 + 8 \kappa^{\prime \, 2} / g^2)$ in the
loose trap limit.  Figure 4 shows $N_{max}$ as a function of $\eps$ for 
various values of $\kappa$.  In general, atom counting will be limited by 
extrinsic linewidth when $16\kappa'^2 > \gpot^2(1-\eps)^2$ and by intrinsic 
linewidth when $16\kappa'^2 < \gpot^2(1-\eps)^2$.

These results demonstrate that atom counting using the transmission 
spectrum is best accomplished within the tight-trap limit.  Certainly, in 
the loose-trap limit, atom counting will be rendered difficult as the 
intrinsic linewidth of the sidebands is increased.  However, several 
questions 
regarding the feasibility of atom counting experiments remain.  First, 
although atom counting by a straightforward measurement of the 
intensity of the transmitted light may be difficult, it is possible 
that the phase of the transmitted light may be less affected by motional 
effects \cite{mabuchi99single}.  Dynamical measurements (possibly using quantum feedback 
techniques) might also yield higher counting limits.  Second, atomic 
cooling techniques could be used in the loose-trap limit to cool the atoms 
into the wells of the optical potential, thereby decreasing the observed 
linewidth \cite{vul,hech,van}.  Finally, the state-dependence of spontaneous emission has not 
yet been taken into account.  
Although the loose-trap regime leads to an
intrinsic linewidth which limits atom counting, it may also suppress the
extrinsic linewidth
as a result of contributions from superluminescence.  
On the other hand, in the 
Lamb-Dicke limit, the atoms are all highly localized, which could lead to 
enhanced spontaneous emission due to cooperative effects.  Future work 
will investigate alternative methods of atom counting and will explore 
complementary techniques of reducing the intrinsic linewidth in atom-cavity
transmission spectra.
\begin{acknowledgements}
We thank Po-Chung Chen for a critical reading of the manuscript. S.L. thanks NSERC for a Postgraduate Scholarship and The Department of Physics of University of California, Berkeley, for a Departmental Fellowship. N.S. thanks the University of California, Berkeley, for a Berkeleyan Fellowship. The work of K.R.B was supported by the Fannie and John Hertz Foundation. The work of D.M.S.K. was supported by the National Science Foundation under Grant No.\ 0130414, the
Sloan Foundation, the David and Lucile Packard Foundation, and the
University of California. KBW thanks the Miller Foundation for Basic Research for a Miller Research Professorship 2002-2003.  The authors' effort was sponsored by the Defense Advanced Research Projects Agency (DARPA) and Air Force Laboratory, Air Force Materiel Command, USAF, under Contract No.\ F30602-01-2-0524.  
\end{acknowledgements}
\pagebreak[4]
\begin{figure}
\includegraphics{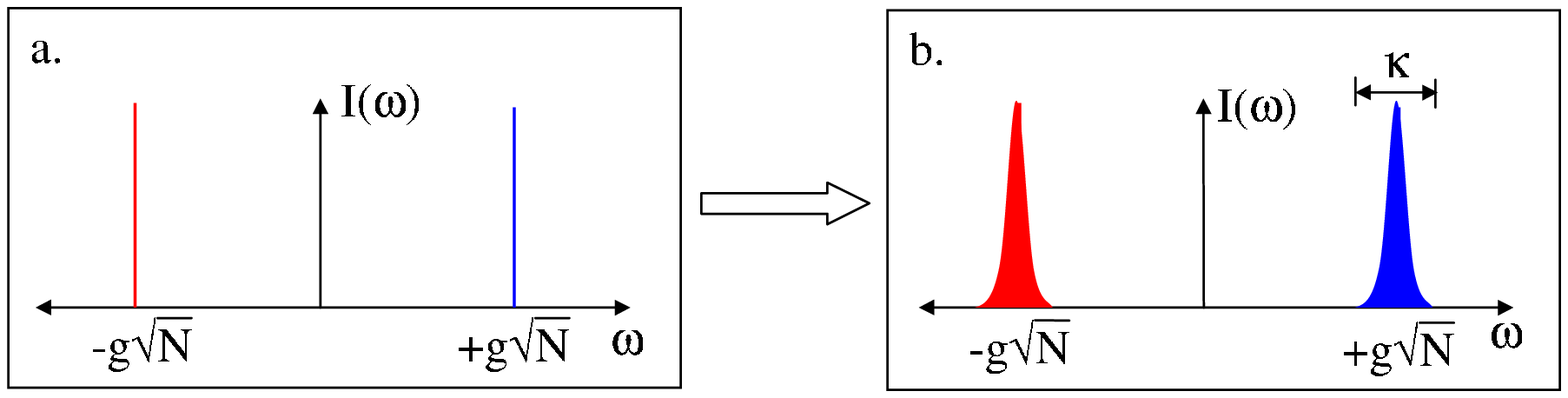} 
\caption{a.  Intrinsic transmission spectrum of atoms-cavity 
system neglecting
spatial dependence of potential and atomic motion.  b.
Transmission spectrum of spatially independent case including
cavity decay.}
\end{figure}

\pagebreak[4]
\begin{figure}
\includegraphics{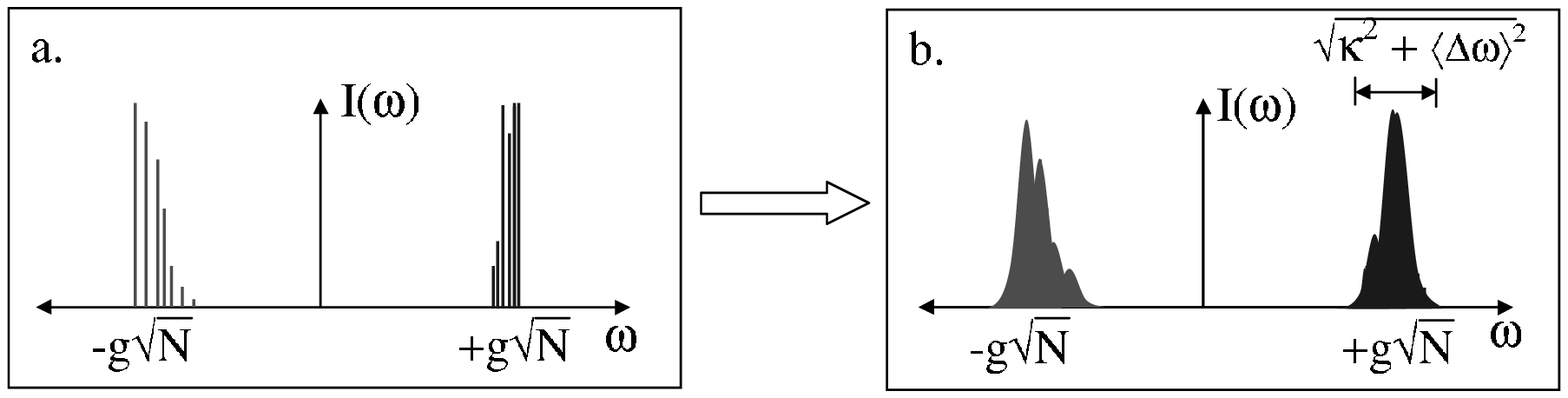}
\caption{a.  Intrinsic transmission spectrum of atoms-cavity 
system including spatial
dependence of potential and atomic motion.  b. Corresponding transmission spectrum including
cavity decay.
}
\end{figure}

\pagebreak[4]
\begin{figure}
\includegraphics{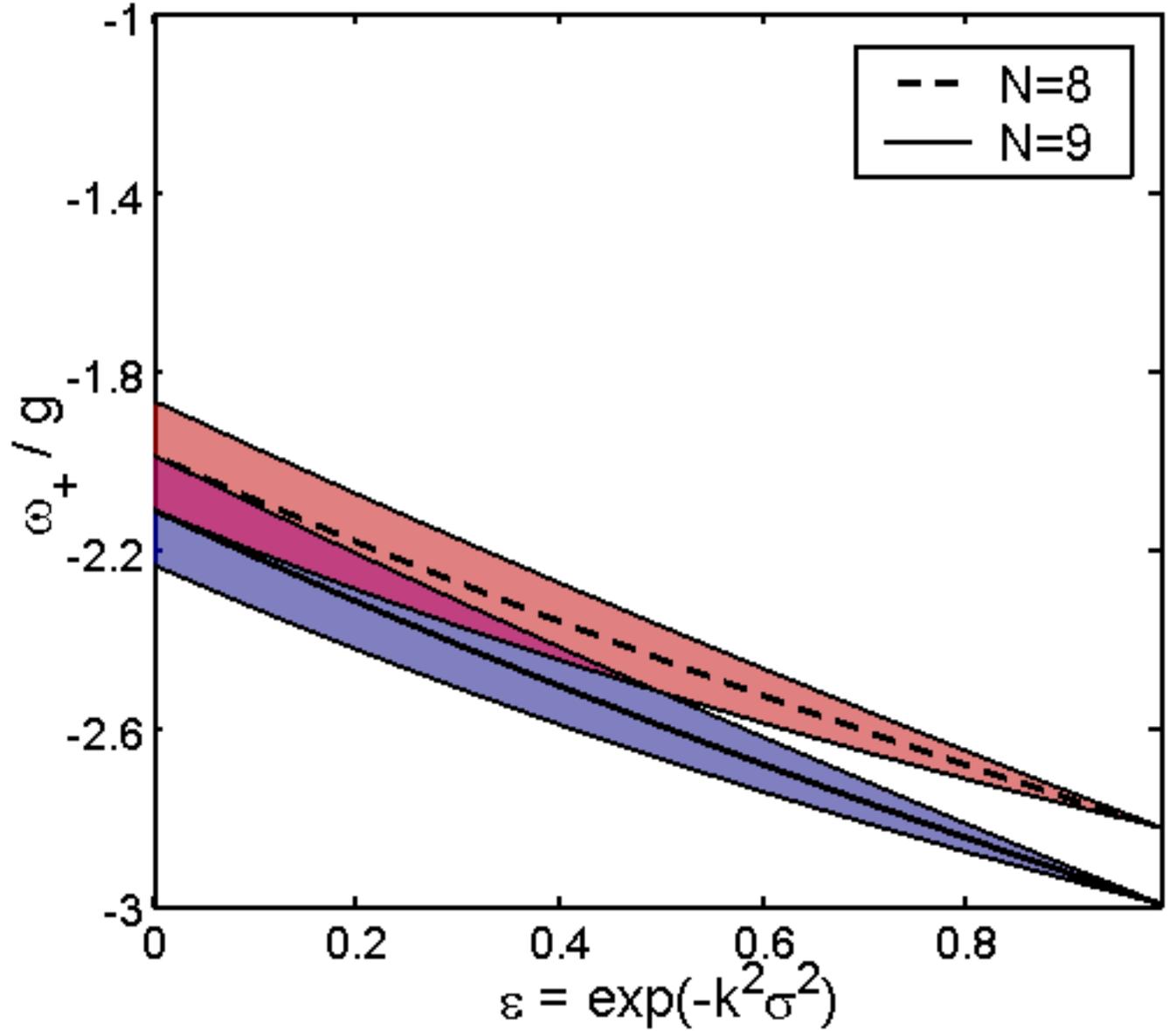}
\caption{
Plot of $\ensavg{\omega_-}$ as a function of the trap tightness $\eps = 
\exp{(-\wvecpot^2\sigma^2)}$ for $N=8$ and $N=9$ and small ratio of atomic recoil energy to vacuum Rabi splitting, $\hbar \wvecpot^2 / 2 m g= 0.01$.  The shaded regions 
indicate the instrinsic 
width of the red sideband, $\pm \sqrt{\ensavg{(\Delta\omega_-)^2}} / 2$.  In the 
tight-trap limit, $N=8$ and $N=9$ can be distinguished.  In the loose-trap 
limit, the intrinsic width of the spectra render determination of atom 
number difficult.
}
\end{figure}

\pagebreak[4]
\begin{figure}
\includegraphics{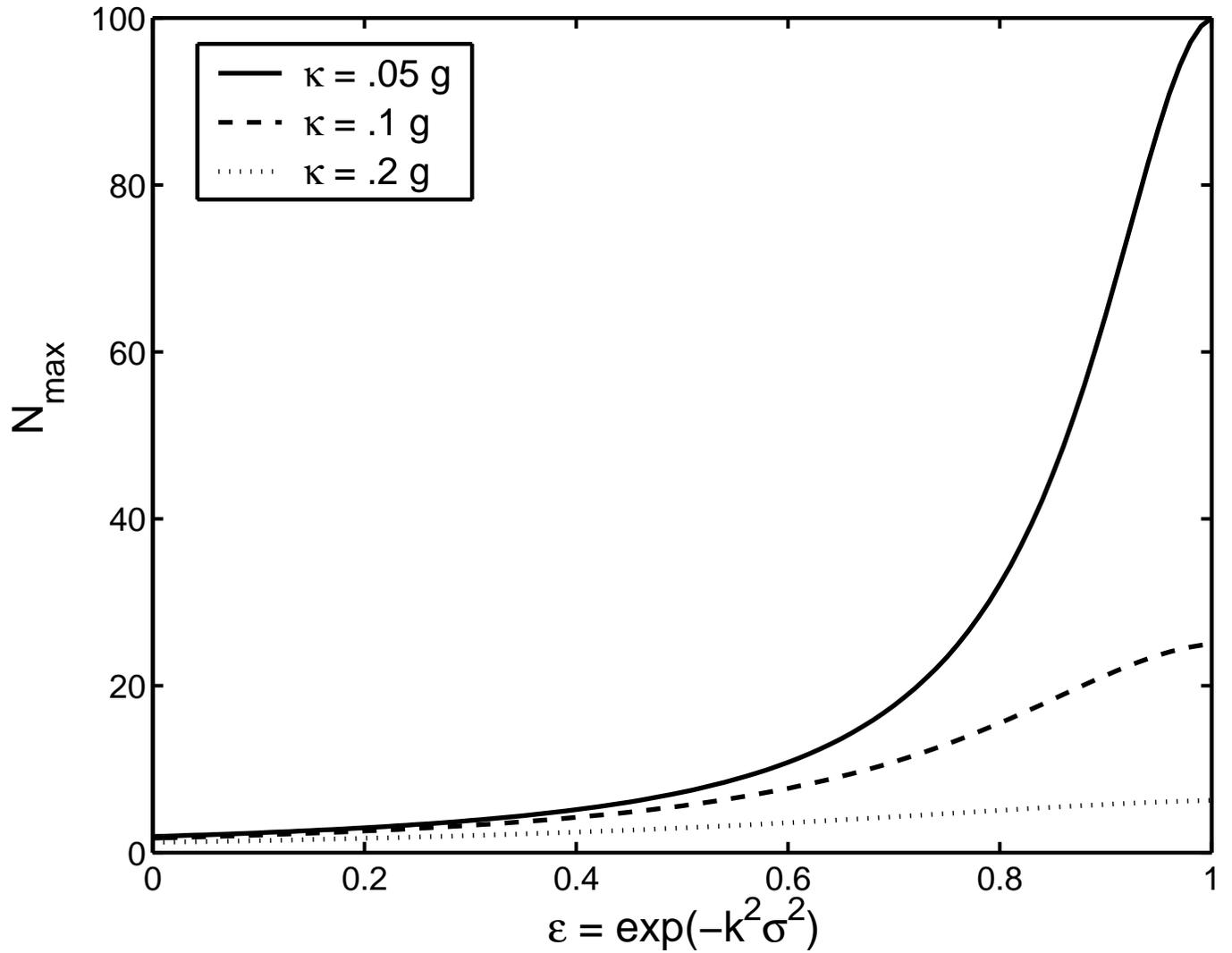}
\caption{
Maximum limit $N_{max}$ on atom counting as a function of trap tightness 
$\eps = \exp{(-\wvecpot^2\sigma^2)}$ for several values of the decay 
parameter $\kappa$.  $\eps \to 0$ corresponds to the 
loose-trap limit while $\eps \to 1$ corresponds to the tight-trap limit.  
Notice that for the infinitely tight trap, atom counting is limited only 
by $\kappa$.
}
\end{figure}

\bibliographystyle{prsty}

\end{document}